\documentclass[aps,prd,onecolumn,groupedaddress,showpacs,nofootinbib,amssymb]{revtex4}
\usepackage[dvips]{graphicx}
\usepackage{amssymb}
\usepackage{amsmath}
\usepackage{graphicx,,color}
\usepackage{amsfonts}
\usepackage{bm}
\usepackage{cancel}
\usepackage{comment}

\newcommand\be{\begin{equation}}
\newcommand\ee{\end{equation}}

\allowdisplaybreaks[4]

\begin{document}

\tolerance=5000

\title{Non-Minimally Coupled Einstein Gauss Bonnet Inflation Phenomenology in View of GW170817}
\author{S.D.~Odintsov,$^{1,2}$\,\thanks{odintsov@ice.cat}
V.K.~Oikonomou,$^{3,4,5}$\,\thanks{v.k.oikonomou1979@gmail.com}F.P.
Fronimos,$^{3}$\,\thanks{fotisfronimos@gmail.com}}
\affiliation{$^{1)}$ ICREA, Passeig Luis Companys, 23, 08010 Barcelona, Spain\\
$^{2)}$ Institute of Space Sciences (IEEC-CSIC) C. Can Magrans
s/n,
08193 Barcelona, Spain\\
$^{3)}$ Department of Physics, Aristotle University of
Thessaloniki, Thessaloniki 54124,
Greece\\
$^{4)}$ Laboratory for Theoretical Cosmology, Tomsk State
University of Control Systems and Radioelectronics, 634050 Tomsk,
Russia (TUSUR)\\
$^{5)}$ Tomsk State Pedagogical University, 634061 Tomsk,
Russia\\}

\tolerance=5000

\begin{abstract}
We study the inflationary phenomenology of a non-minimally coupled
Einstein Gauss-Bonnet gravity theory, in the presence of a scalar
potential, under the condition that the gravitational wave speed
of the primordial gravitational waves is equal to unity, that is
$c_T^2=1$, in natural units. The equations of motion, which are
derived directly from the gravitational action, form a system of
differential equations with respect to Hubble's parameter and the
inflaton field which are very complicated and cannot be solved
analytically, even in the minimal coupling case. In this paper, we
present a variety of different approximations which could be used,
along with the constraint $c_T^2=1$, in order to produce an
inflationary phenomenology compatible with recent observations.
All the different approaches are able to lead to viable results if
the model coupling functions obey simple relations, however,
different approaches contain different approximations which must
be obeyed during the first horizon crossing, in order for the
model to be rendered correct. Models which may lead to a
non-viable phenomenology are presented as well in order to
understand better the inner framework of this theory. Furthermore,
since the velocity of the gravitational waves is set equal to
$c_T^2=1$, as stated by the striking event of GW170817 recently,
the non-minimal coupling function, the Gauss-Bonnet scalar
coupling and the scalar potential are related to each other. Here,
we shall assume no particular form of the scalar potential and we
choose freely the scalar functions coupled to the Ricci scalar and
the Gauss-Bonnet invariant. Certain models are also studied in
order to assess the phenomenological validity of the theory, but
we need to note that all approximations must hold true in order
for a particular model to be valid. Finally, even though each
possible approach assumes different approximations, we summarize
them in the last section for the sake of completeness.
\end{abstract}


\maketitle

\section{Introduction}

The recent years have been proved to be outstanding for physicists
and in particular, theoretical cosmologists. The most striking
observation of the previous century that in a sense, has proved to
us the incomplete perception of physicists, when it comes to
understanding the secrets of the cosmos, was the realization that
the Universe we live in does not only expand, but it expands with
an accelerating rate \cite{Riess:1998cb}. As striking as it may
sound, it is not in fact the only era in which the Universe may
have exhibited an accelerating expansion. The inflationary era,
which occurred moments after the Big Bang, also describes an
accelerating expansion.

The era of inflation is very interesting and peculiar. Inflation
states that the Universe experienced a drastic expansion in a tiny
time interval, moments after the Big Bang occurred and before
other important events such as the electroweak baryogenesis. From
that moment, it is stated that the Universe throughout the years
has not managed to expand as much as during the inflationary era.
Furthermore, it is regarded as an essential tool which promises to
shine light towards many problems which to this day remain
unsolved. Inflation provides a possible explanation for the
observed flatness of our Universe, the absence of magnetic
monopoles which are predicted in many theories concerning high
energy physics and the cosmological perturbations in matter and
radiation which are currently observed. For instance the absence
of magnetic monopoles can be attributed to the exponential
expansion of the Universe, which led to a decrease in their
average density at large scales.

Inflation however, even if it is considered a very important event
in cosmology and high energy physics, it does not specify the
framework of gravity that can produce such an era. In other words,
it can be realized even if the theory of gravity is not that of
Einstein's but is a different, modified theory
\cite{Nojiri:2017ncd,Nojiri:2010wj,Nojiri:2006ri,Capozziello:2011et,Capozziello:2010zz,delaCruzDombriz:2012xy,Olmo:2011uz}.
In common literature, there exists a plethora of models for
modified gravity theories which manage to explain many
observations. Such theories could also produce a plausible
scenario for the aforementioned era of inflation. These possible
scenarios may be endless, but in recent years, the precise
observations have managed to rule out many promising models. The
main observation which is the factor that decides the validity of
a modified theory of gravity, and of gravity in general, is the so
called Cosmic Microwave Background (CMB). This radiation is
diffused in the observable Universe and contains information which
was encoded to photons during the first horizon crossing, freezed
until the last scattering surface. By studying the CMB, we can
discard theories which are unable to be compatible with the
observations. Some of the information which can be extracted from
the CMB is quantified by the spectral index of the primordial
scalar perturbations and the tensor-to-scalar ratio.

There exist also many modified theories of gravity, which can
manage to survive the precision tests of the CMB but still be
intrinsically unrealistic. Nowadays, these theories can be tested
and even discarded if deemed necessary, by studying strong sources
of gravity. The era of multi-messenger astronomy is before us and
provides the appropriate data to distinguish the realistic models
from the unrealistic ones. Today, we are able to study cosmic
events in two separate ways, firstly by witnessing the
electromagnetic phenomenon, as was custom for the section of
astronomy during the last century, and secondly by examining the
gravitational waves emitted by strong sources of gravity. This new
way of perceiving the Universe has led theorists to accept a
fascinating result, that the gravitational waves, or in other
words the perturbations in the metric, propagate through spacetime
with their velocity being equal to that of light. For theorists
which study modified theories of gravity, this is a striking
result, since theoretical frameworks which propose a different
velocity which in fact deviates from the speed of light, must be
discarded. In Ref.\cite{Ezquiaga:2017ekz} a list of such theories
is presented in detail. This result seems to be indicative of the
fact that nature will always find a way to convince us whether a
model for describing it is actually correct or otherwise false.

A theory belonging to the previous category is the Einstein
Gauss-Bonnet gravity theory
\cite{Hwang:2005hb,Nojiri:2006je,Cognola:2006sp,Nojiri:2005vv,Nojiri:2005jg,Satoh:2007gn,Bamba:2014zoa,Yi:2018gse,Guo:2009uk,Guo:2010jr,Jiang:2013gza,Kanti:2015pda,vandeBruck:2017voa,Kanti:1998jd,Kawai:1999pw,Nozari:2017rta,Chakraborty:2018scm,Odintsov:2018zhw,Kawai:1998ab,Yi:2018dhl,vandeBruck:2016xvt,Kleihaus:2019rbg,Bakopoulos:2019tvc,Maeda:2011zn,Bakopoulos:2020dfg,Ai:2020peo,Odintsov:2020sqy,Odintsov:2020zkl,Easther:1996yd,Antoniadis:1993jc}
which, as a matter of fact, is the one we shall work with in this
paper. In this type of theories however, the gravitational waves
do not have a fixed value for their velocity and therefore can be
set equal to the velocity of light, by forcing the
Gauss-Bonnet coupling scalar function to obey a specific relation.
This is a powerful characteristic since now these theories can
survive the test of recent observations, such was the GW170817
\cite{GBM:2017lvd}. This particular event established the term
multi-messenger astronomy which we referred to previously and had
a great impact on not only Cosmology, but also Nuclear Physics, as
it provided also a mechanism for the creation of heavy nuclei.

In the present paper we extend the formalism of a recent previous
work of ours \cite{Odintsov:2020sqy}, in order to study
Einstein-Gauss-Bonnet gravity inflationary phenomenology, in the
presence of a non-minimal coupling of the scalar field to the
Ricci scalar. This is a different category of theories which
contain a function of the scalar field, coupled with the Ricci
scalar, which specifies the average curvature in a region. This
coupled term leads to extra geometric terms in the gravitational
equations of motion, or in other words it may lead to new physics,
since the theory does not have an Einstein frame counterpart
theory. Moreover, it can lead to simplifications or even viability
of certain models which would otherwise had to be rendered as
physically unrealistic. Our aim is to present several models of
non-minimally coupled Einstein-Gauss-Bonnet theory, in the
presence of a scalar potential too, and confront their
inflationary phenomenology with the observational data.

This paper is outlined as follows: In the first section, we
present the theoretical framework of the non-minimally coupled
Einstein-Gauss-Bonnet theory, in the presence of a scalar
potential, and we also demonstrate the constraints imposed by the
condition $c_T^2=1$. It is worth mentioning that even though the
GW179817 event does not refer to the inflationary era, this constraint is imposed 
in the present paper in order to obtain a massless primordial graviton from 
the perspective of particle physics. Then we introduce the slow-roll indices and
assume the least possible approximations in order to make the
system of the equations of motion elegant and solvable. In the
following sections, we present the form of the observed quantities
according to this particular framework and develop certain models
for a number of possible approaches one could follow in order to
solve the aforementioned system of equations of motion and
thoroughly study the validity of the approximations made. Finally,
in the last section we present all the possible approaches one
might follow in order to solve the equations of motion and derive
acceptable results. Every possible approach is accompanied by
necessary approximations that must apply during the first horizon
crossing, in order for the model to be called viable.

Before we begin our study, it is worth specifying the cosmological
background we shall assume. In this paper, we shall assume a flat
Friedman-Robertson-Walker (FRW) metric corresponding to a line
element,
\begin{equation}
\label{metric} \centering ds^2-dt^2+a^2(t)\sum_{i=1}^{3}{(dx^i)^2}\,,
\end{equation}
where $a(t)$ denotes the scale factor.

\section{Essential Features of GW170817 non-minimally Coupled Einstein-Gauss-Bonnet Gravity and Inflationary Phenomenology}

The starting point of our study is obviously the gravitational
action, since all the information about the Universe at the era of
inflation is contained in it. Let us assume that the action is
defined as,
\begin{equation}
\label{action} \centering
\mathcal{S}=\int{d^4x\sqrt{-g}\left(\frac{h(\phi)R}{2\kappa^2}-X-V(\phi)+\mathcal{L}_c\right)}\,
,
\end{equation}
where $g$ is the determinant of the metric tensor,
$\kappa=\frac{1}{M_P}$ is a constant proportional to the reduced
Planck mass, $h(\phi)$ is a dimensionless scalar function coupled
to the Ricci scalar $R$, $X$ is the kinetic term designated as
$X=\frac{1}{2}\omega g^{\mu\nu}\partial_\mu\phi\partial_\nu\phi$,
$V$ is the scalar potential and finally, $\mathcal{L}_c$ denotes
the string corrections which are specified as
$\mathcal{L}_c=-\frac{1}{2}\xi(\phi)\mathcal{G}$. Here,
$\xi(\phi)$ denotes the Gauss-Bonnet coupling scalar function and
$\mathcal{G}$ signifies the Gauss-Bonnet invariant defined as
$\mathcal{G}=R^2-R_{\alpha\beta}R^{\alpha\beta}+R_{\alpha\beta\gamma\delta}R^{\alpha\beta\gamma\delta}$,
where $R_{\alpha\beta}$ and $R_{\alpha\beta\gamma\delta}$ are the
Ricci and Riemann tensor respectively. Since the line element
corresponds to that of a flat Friedman-Robertson-Walker spacetime,
then certain terms in the gravitational action are simplified.
Specifically, the kinetic term is now written as
$X=-\frac{1}{2}\omega\dot\phi^2$ by simply assuming that the
scalar field is homogeneous and additionally, the Gauss-Bonnet
scalar is written as $\mathcal{G}=24H^2(\dot H+H^2)$. Here, the
``dot'' represents differentiation with respect to the cosmic
time. Last but not least, we mention that even though we shall
work with a canonical kinetic term, the $\omega$ parameter will be
kept undefined for the moment, instead of being replaced with
$\omega=1$ in order to have the phantom case $\omega=-1$ available
too.

As mentioned before, the gravitational action contains all the
information available, so the equations of motion, which are
necessary in order to describe the dynamics of inflation, can be
extracted from Eq. (\ref{action}) by implementing the variation
principle. Consequently, the equations of motion are written as
\begin{equation}
\label{motion1} \centering
\frac{3hH^2}{\kappa^2}=\frac{1}{2}\omega\dot\phi^2+V-\frac{3H\dot
h}{\kappa^2}+12\dot\xi H^3 \, ,
\end{equation}
\begin{equation}
\label{motion2} \centering -\frac{2h\dot
H}{\kappa^2}=\omega\dot\phi^2-\frac{H\dot h}{\kappa^2}-8\dot\xi
H\dot H+\frac{\ddot\phi
h'+h''\dot\phi^2}{\kappa^2}-4H^2(\ddot\xi-\dot\xi H)\, ,
\end{equation}
\begin{equation}
\label{motion3} \centering
\ddot\phi+3H\dot\phi+\frac{1}{\omega}\left(V'-\frac{Rh'}{2\kappa^2}+12\xi'H^2(\dot
H+H^2)\right)=0\, ,
\end{equation}
where the ``prime'' denotes differentiation with respect to the
scalar field $\phi$. Solving this particular system of
differential equations requires finding an analytical expression
for Hubble's parameter and the scalar field $\phi$, which should
give us a complete description of the inflationary era.
Unfortunately, these equations are very complicated and the system
cannot be solved analytically. The solution may be extracted, only
if certain approximations are made which facilitate our study and
in fact make the system solvable. Before we proceed with the
approximations, we shall impose a strong constraint on the
velocity of the gravitational waves in order to achieve
compatibility with the recent GW170817 observation.

Gravitational waves are perturbations in the metric which travel
through spacetime with the speed of light  \cite{GBM:2017lvd}, as
it was ascertained recently. However, theories which contain
string corrections lead to an expression for their cosmological
tensor perturbations velocity, which in fact deviates from the
speed of light. The general expression in a cosmological context
is,
\begin{equation}
\centering c_T^2=1-\frac{Q_f}{2Q_t}\, ,
\end{equation}
where $Q_f=8(\ddot\xi-H\dot\xi)$, $Q_t=F+\frac{Q_b}{2}$,
$Q_b=-8\dot\xi H$ and $F=\frac{h}{\kappa^2}$. If the gravitons
are massless during and after the inflationary era, if we demand
that $Q_f=0$, meaning that $\ddot\xi=H\dot\xi$, we get $c_T^2=1$.
As a result, Eq. (\ref{motion2}) is greatly simplified and leads
us one step closer to finding simplified solutions for the
inflationary era. Apart from solving the problem with the velocity
of gravitational waves, a simple differential equation is derived,
which reads,
\begin{equation}
\centering \ddot\xi=H\dot\xi \, ,
\end{equation}
Instead of solving this differential equation with respect to the
cosmic time, like was done in Ref. \cite{Odintsov:2019clh}, we can
modify it properly and extract a deeper connection between the
scalar field $\phi$ and the Gauss-Bonnet coupling scalar function
$\xi$ \cite{Odintsov:2020sqy}. Since
$\frac{d}{dt}=\dot\phi\frac{d}{d\phi}$, the differential equation
can be rewritten as,
\begin{equation}
\centering \dot\phi^2\xi''+\xi'\ddot\phi=H\xi'\dot\phi\, ,
\end{equation}
This equation can be simplified if the slow-roll approximation is
considered. Let us assume that $\ddot\phi\ll H\dot\phi$. Hence,
\begin{equation}
\label{dotphi} \centering \dot\phi\simeq\frac{H\xi'}{\xi''}\, ,
\end{equation}
This is a simple correlation between the derivative of the scalar
field and the first two derivatives of the Gauss-Bonnet coupling
scalar function. It can be used in Eqs.
(\ref{motion1})-(\ref{motion3}), in order to replace the
derivative of scalar field. We can further simplify the equations
of motion, by using the slow-roll approximation, which is
essential to inflationary phenomenology. Assuming in addition that
the kinetic term is insignificant compared to the scalar potential
and also Hubble's derivative is also lesser than Hubble's
parameter squared, that is,
\begin{align}
\label{slowrollapprox} \centering \dot H&\ll H^2&
\frac{1}{2}\omega\dot\phi^2&\ll V&\ddot\phi&\ll H\dot\phi\, ,
\end{align}
then the equations of motion can be greatly simplified as shown
below,
\begin{equation}
\label{motion4} \centering
\frac{3hH^2}{\kappa^2}=V-\frac{3H^2h'}{\kappa^2}\frac{\xi'}{\xi''}-12\frac{\xi'^2}{\xi''}H^4\,
,
\end{equation}
\begin{equation}
\label{motion5} \centering -\frac{2h\dot H}{k^2}=\omega
H^2\left(\frac{\xi'}{\xi''}\right)^2-\frac{H^2h'}{\kappa^2}\frac{\xi'}{\xi''}-8\frac{\xi'^2}{\xi''}H^2\dot
H+\frac{h''H^2}{\kappa^2}\left(\frac{\xi'}{\xi''}\right)^2\, ,
\end{equation}
\begin{equation}
\label{motion6} \centering
V'+3H^2\left(\omega\frac{\xi'}{\xi''}-2\frac{h'}{\kappa^2}+4\xi'H^2\right)=0\,
,
\end{equation}
These are the gravitational equations of motion simplified due to
the assumption of the slow-roll approximation, and due to the fact
that we assumed $c_T^2=1$ for the primordial tensor perturbations.
However, even simplified to this form, the system remains
unsolvable. More approximations are needed in order to make the
equations solvable and examine the viability of a model. The
appealing characteristic of our results is that Hubble's
derivative is written proportionally to Hubble's parameter. Hence,
Eq. (\ref{motion5}) describes as we shall see, the slow-roll index
$\epsilon_1$, which is a powerful relation since different
assumptions could lead to different approaches to the problem and
therefore different results.

The dynamics of inflation can be described by studying the
slow-roll indices \cite{Hwang:2005hb}, which are defined as,
\begin{align}
\centering \epsilon_1&=\pm \frac{\dot H}{H^2}
&\epsilon_2&=\frac{\ddot\phi}{H\dot\phi} &\epsilon_3&=\frac{\dot
F}{2HF} & \epsilon_4&=\frac{\dot E}{2HE} &\epsilon_5&=\frac{\dot
F+Q_a}{2HQ_t} & \epsilon_6&=\frac{\dot Q_t}{2HQ_t}\, ,
\end{align}
where $E=F\left(\omega+\frac{3(\dot
F+Q_a)^2}{2\dot\phi^2Q_t}\right)$ and $Q_a=-4\dot\xi H^2$. The
sign of slow-roll index $\epsilon_1$ seems arbitrary but it is
convenient to assume either the positive or the negative value in
certain cases. For the purposes of this paper we shall take the
positive value though. Furthermore, the expression of the index,
as mentioned before, is depending on the approximations which will
be implemented in Eq. (\ref{motion5}) so we will refrain from
writing an analytic form at this point. In the following models,
we shall specify the sign of $\epsilon_1$ before commencing with
the results. However it is worth writing the forms of the rest of
the slow-roll indices,
\begin{equation}
\centering \epsilon_2\simeq 1+\frac{\dot
H}{H^2}-\frac{\xi'\xi'''}{\xi'^2}\, ,
\end{equation}
\begin{equation}
\centering
\epsilon_3\simeq\frac{1}{2}\frac{\xi'}{\xi''}\frac{h'}{h}\, ,
\end{equation}
\begin{equation}
\centering
\epsilon_4\simeq\epsilon_3+\frac{1}{2}\frac{\xi'}{\xi''}\frac{P'}{P}\,
,
\end{equation}
\begin{equation}
\centering
\epsilon_5\simeq\frac{1}{2Q_t}\frac{\xi'}{\xi''}\left(\frac{h'}{\kappa^2}-4\xi'H^2\right)\,
,
\end{equation}
\begin{equation}
\centering
\epsilon_6\simeq\frac{H}{2Q_t}\frac{\xi'}{\xi''}\left(2(1+\frac{\dot
H}{H^2})+H\frac{\xi'^2\xi'''}{\xi''^2}\right)\, ,
\end{equation}
where we introduced $P=\frac{E}{F}$ for convenience. As we can see
see, indices $\epsilon_2$ and $\epsilon_6$ are connected with
$\epsilon_1$ and in addition, index $\epsilon_4$ with
$\epsilon_3$. That does not mean however that indices $\epsilon_3$
and $\epsilon_4$ are equivalent.

Finally, we examine the form of the $e$-foldings number $N$, which
is of fundamental importance in our study. By definition, the
$e$-foldings number is written as $N=\int_{t_i}^{t_f}{Hdt}$ where
$t_i$ and $t_f$ signify the initial and final moment of inflation,
or to put it simply, the difference $t_f-t_i$ signifies the
duration of inflation. However, using Eq. (\ref{dotphi}), one can
alter the variable and work solely with the scalar field $\phi$.
As a result, the formula for the $e$-foldings number is altered as
shown below,
\begin{equation}
\label{efolds} \centering
N=\int_{\phi_i}^{\phi_f}{\frac{\xi''}{\xi'}d\phi}\, ,
\end{equation}
This form implies that the $e$-foldings number is strongly
dependent on the choice of the function $\xi(\phi)$, so by
choosing a simple coupling scalar function $\xi(\phi)$, or one
with appropriate characteristics, could yield in principle a
simple phenomenology. In the following section we shall
appropriately choose both coupling functions and study the
phenomenology of the non-minimally coupled Einstein-Gauss-Bonnet
model, and how viability can be achieved imposing certain
necessary approximations in the equations of motion. It will also
be shown that even though certain approaches seem fascinating, in
the end do not result to viable models.

\section{Models of non-minimally Coupled Einstein-Gauss-Bonnet
Gravity and Compatibility with Planck Data}

Beginning this paper, we wrote down the gravitational action
(\ref{action}), which was the starting point in deriving the
equations of motion. This equation has certain unspecified
functions, mainly the coupling functions $h(\phi)$ and
$\xi(\phi)$, along with the scalar potential $V(\phi)$.
Consequently, in order to derive the expression of Hubble's
parameter, these functions must be determined. However, since the
constraint in the velocity of the gravitational waves was imposed,
the scalar potential depends on the other two freely chosen
functions. In fact, the third equation of motion connects both
coupling functions to the scalar potential. Thus, when specifying
the coupling functions, the potential cannot take an arbitrarily
chosen form, but must obey Eq. (\ref{motion6}). In the following
models, we shall assume that the scalar potential obeys a more
simplified differential equation, which is,
\begin{equation}
\label{Vdifeq1} \centering
V'+3H^2\left(\omega\frac{\xi'}{\xi''}-2\frac{h'}{\kappa^2}\right)=0\,
,
\end{equation}
This assumption is not necessary but it is convenient, since a
more manageable potential may be derived, but we note that the
assumption $\xi'H^4\ll V'$, along with the slow-roll
approximations (\ref{slowrollapprox}), must hold true in order for
the model to be viable. These assumptions, in addition to the rest
which shall make hereafter, will be validated if these hold true
at the end of each model.

In order to ascertain the validity of a model, the results which
the model produces must be confronted to the recent Planck
observational data \cite{Akrami:2018odb}. In the following models,
we shall derive the values for the quantities, namely the spectral
index of primordial curvature perturbations $n_S$, the
tensor-to-scalar-ratio $r$ and finally, the tensor spectral index
$n_T$  \cite{Hwang:2005hb}. These quantities are connected with
the slow-roll indices introduced previously, as shown below,
\begin{align}
\label{results} \centering n_S&=1+2\frac{2\frac{\dot
H}{H^2}-\epsilon_2+\epsilon_3-\epsilon_4}{1+\frac{\dot H}{H^2}} &
n_T&=2\frac{\frac{\dot H}{H^2}-\epsilon_6}{1+\frac{\dot H}{H^2}}
&r&=16\left|\left(\frac{Q_e}{4HF}+\frac{\dot
H}{H^2}-\epsilon_3\right)\frac{Fc_A^3}{Q_t}\right|\, ,
\end{align}
where $c_A$ is the sound wave velocity defined as,
\begin{equation}
\centering c_A^2=1+\frac{Q_e(\dot
F+Q_a)}{2\omega\dot\phi^2Q_t+3(\dot F+Q_a)^2}\, ,
\end{equation}
Since the sign of slow-roll index $\epsilon_1$ has not been
specified yet, it was deemed necessary to write the ratio
$\frac{\dot H}{H^2}$. Moreover, spectral index $n_T$ has not been
experimentally verified to date, since no B-modes have been
observed so far in the CMB. However we shall call it an observable
quantity and make a prediction for its value for each model.
According to the recent Planck 2018 collaboration
\cite{Akrami:2018odb}, the rest observed quantities have the
following values,
\begin{align}
\centering n_S&=0.9649\pm0.0042& r&<0.064\, ,
\end{align}
These values can be theoretically evaluated by inserting the
wavenumber $k$ during the first horizon crossing as a input. Here,
we have followed a different approach. Instead of using
wavenumbers, we study the era of inflation using the inflaton
field $\phi$. As a result, the previous values can be calculated
by inserting the value of the scalar field at the start of
inflation as an input. However, in order to evaluate the initial
value of the scalar field during this era, one must find first the
final value of the scalar field. This value is derived easily by
equating slow-roll index $|\epsilon_1|$ with 1. This is exactly
why this slow-roll index has not been properly designated so far,
because different approximations yield different forms and
therefore different expressions for the final value of the scalar
field. In the following models, we shall firstly designate both
coupling scalar functions, specify the sign of slow-roll index
$\epsilon_1$, make certain assumptions, derive results, compare
them with the observations and assess the validity of the
assumptions made in each model separately. Before we proceed
however, it is worth writing the analytic form of the auxiliary
functions $Q_a$, $Q_b$, $Q_e$ and $Q_t$, which are,
\begin{equation}
\centering Q_a\simeq-4\frac{\xi'^2}{\xi''}H^3\, ,
\end{equation}
\begin{equation}
\centering Q_b\simeq-8\frac{\xi'^2}{\xi''}H^2\, ,
\end{equation}
\begin{equation} \centering
Q_e\simeq-16\frac{\xi'^2}{\xi''}H\dot H\, ,
\end{equation}
\begin{equation}
\centering
Q_t\simeq\frac{h}{\kappa^2}-4\frac{\xi'^2}{\xi''}H^2\, .
\end{equation}
In the following subsections, we shall examine the phenomenology
of several models by specifying the functional forms of the
non-minimal coupling scalar function $h(\phi)$ and of the
Gauss-Bonnet coupling function $\xi(\phi )$.

\subsection{A Model with Power-Law $h(\phi)$ and Exponential $\xi(\phi )$ Functions}

This model is supposedly one of the easiest imaginable. The
coupling scalar functions are defined as,
\begin{equation}
\label{h1} \centering h(\phi)=\Lambda_1(\kappa\phi)^{n_1}\, ,
\end{equation}
\begin{equation}
\label{xi1} \centering \xi(\phi)=\lambda_1 e^{\gamma_1\kappa\phi}\, ,
\end{equation}
The reasons behind choosing such functions is because the first is
a very simple case of power-law, while the latter has an appealing
characteristic. In the previous equations, the coupling functions
appear in ratios of $\xi'/\xi''$, $h'/h$ and $h''/h$, so it is only
reasonable to try and use functions which simplify greatly these
ratios. This is exactly why the exponential function was chosen
along with the power-law, since,
\begin{align}
\centering h'&=\frac{n_1}{\phi}h
&h''&=\frac{n_1(n_1-1)}{\phi^2}h&\xi''&=\kappa\gamma_1\xi'\, ,
\end{align}
Now, all that remains is to specify the forms of Hubble's
parameters. In this models, we shall assume that,
\begin{equation}
\label{motion7} \centering H^2\simeq\frac{k^2V}{3h}\, ,
\end{equation}
\begin{equation}
\label{motion8} \centering \dot
H\simeq\frac{H^2}{2}\left(\frac{h'}{h}\frac{\xi'}{\xi''}-\frac{h''}{h}\left(\frac{\xi'}{\xi''}\right)^2\right)\,
,
\end{equation}
These relations are derived from Eqs. (\ref{motion4}) and
(\ref{motion5}) respectively. In the end, we shall examine whether
the assumptions made hold true.

\begin{figure}[h!]
\centering
\includegraphics[width=17pc]{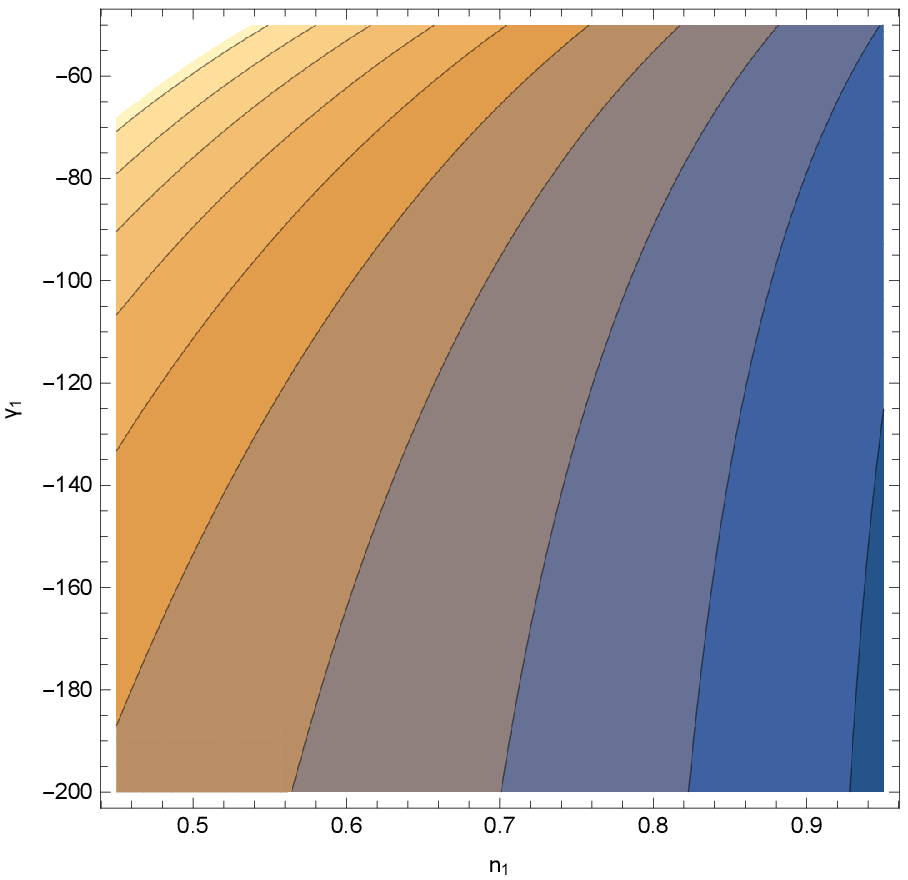}
\includegraphics[width=3pc]{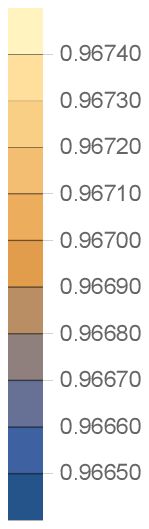}
\includegraphics[width=17pc]{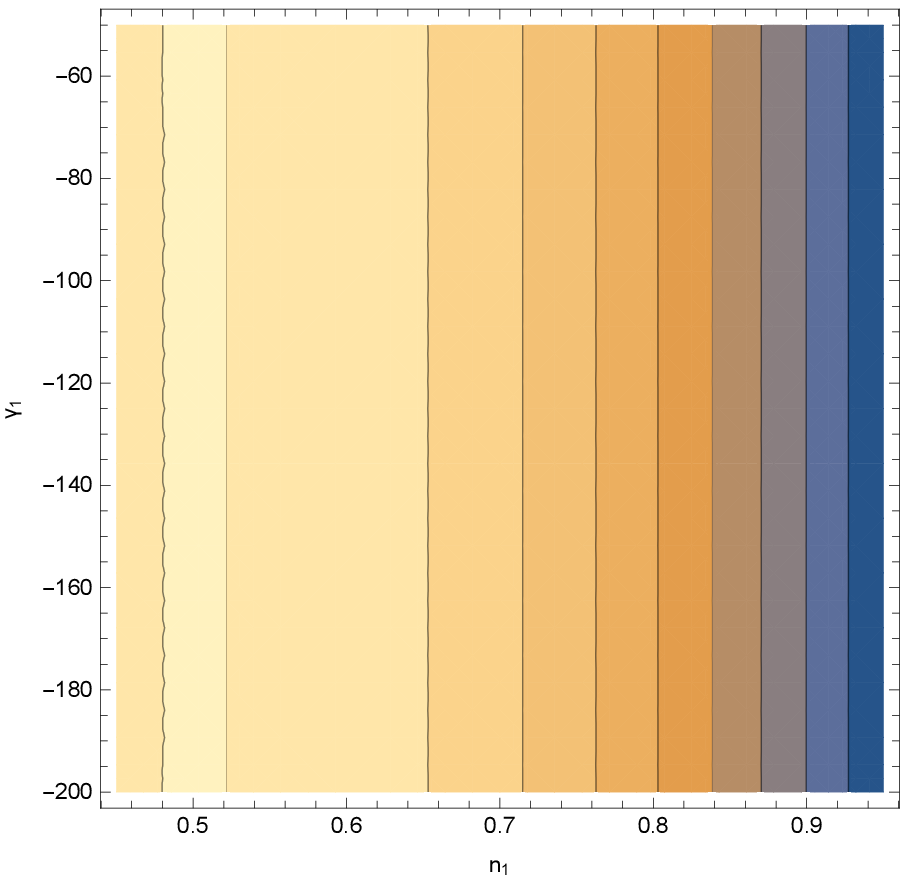}
\includegraphics[width=3pc]{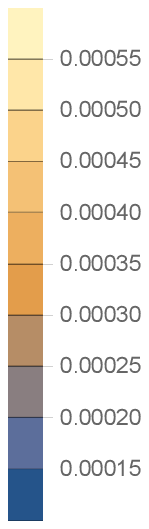}
\caption{Parametric plot of spectral index $n_S$ (left) and
tensor-to-scalar ratio $r$ (right) depending on exponent $n_1$ and
$\gamma_1$, ranging from [0.45, 0.95] and [-200,-50] respectively. It
can be inferred that there exists a variety of possible values
for the free parameters which lead to compatible observed indices. }
\label{plot1}
\end{figure}

From the expression of the Hubble's derivative, we see that it is
appropriate to choose the positive sign for index $\epsilon_1$,
that is, $\epsilon_1=\frac{\dot H}{H^2}$. Thus, from Eq.
(\ref{motion8}), it can be inferred that the slow-roll index
$\epsilon_1$ and subsequently the form of the final value of the
scalar field depend strongly on the ratios of the coupling
functions with their derivatives respectively. Once again, the
selection of a power-law and an exponential function seems
appropriate since they  lead to simplified form. Before we proceed
with the expressions of the slow-roll indices, let us first derive
the expression for the scalar potential. From Eq. (\ref{Vdifeq1}),
\begin{equation}
\label{pot1} \centering
V(\phi)=V_1(\kappa\phi)^{2n_1}Exp\left(\alpha_1(\kappa\phi)^{1-n_1}\right)\, ,
\end{equation}
where $\alpha_1=\frac{\omega}{\gamma_1\Lambda_1(n_1-1)}$ and $V_1$ is a
constant with mass dimensions [m]$^4$ . As a result, the
slow-roll indices can be evaluated, in certain cases having simple
expressions, as shown below,
\begin{equation}
\label{index1A} \centering
\epsilon_1\simeq\frac{n_1}{2\gamma_1\kappa\phi}\left(1-\frac{n_1-1}{\gamma_1\kappa\phi}\right)\,
,
\end{equation}
\begin{equation}
\label{index2A} \centering
\epsilon_2\simeq\frac{n_1}{2\gamma_1\kappa\phi}\left(1-\frac{n_1-1}{\gamma_1\kappa\phi}\right)\,
,
\end{equation}
\begin{equation}
\label{index3A} \centering
\epsilon_3\simeq\frac{n_1}{2\gamma_1\kappa\phi}\, ,
\end{equation}
\begin{equation}
\label{index5A} \centering \epsilon_5\simeq\frac{3 \Lambda_1 ^2 n_1
(\kappa  \phi )^{2 n_1}-4 \gamma_1 \kappa\phi \lambda_1 \kappa^4
V(\phi ) e^{\gamma_1  \kappa  \phi }}{6 \gamma_1  \kappa  \Lambda_1 ^2
\phi  (\kappa  \phi )^{2 n_1}-8 \gamma_1 \kappa\phi \lambda_1
\kappa^4V(\phi ) e^{\gamma_1  \kappa  \phi }}\, ,
\end{equation}
\begin{equation}
\label{index6A} \centering \epsilon_6\simeq\frac{4 \lambda_1  n_1
\kappa^4V(\phi ) e^{\gamma_1  \kappa  \phi }-4 \lambda_1
\kappa\phi  e^{\gamma_1  \kappa  \phi } \kappa^3V'(\phi )-4 \gamma_1
\kappa\phi \lambda_1 \kappa^4V(\phi ) e^{\gamma_1  \kappa  \phi
}+3 \Lambda_1 ^2 n_1 (\kappa  \phi )^{2 n_1}}{6 \gamma_1  \kappa\phi
\Lambda_1 ^2   (\kappa  \phi )^{2 n_1}-8 \gamma_1 \kappa\phi
\lambda_1  \kappa^4V(\phi ) e^{\gamma_1  \kappa  \phi }}\, ,
\end{equation}
It is obvious that the first three indices have simple functional
forms, while the rest are more complicated, especially the index
$\epsilon_4$, which is omitted due to this reason. Let us now
continue with the evaluation of the necessary values of the
inflaton field. Firstly, as mentioned before, the final value of
the scalar field can be extracted by equating index $\epsilon_1$
to unity. As a result, the final value of the scalar field has the
following form,
\begin{equation}
\label{scalarfA} \centering
\phi_f=\frac{n_1\gamma_1+\sqrt{\gamma_1^2n_1(8-7n_1)}}{4\gamma_1^2\kappa}\, .
\end{equation}
Utilizing the form of the $e$-foldings number in Eq.
(\ref{efolds}), the initial value of the scalar field is extracted
and subsequently the observed quantities. The initial value reads,
\begin{equation}
\label{scalariA} \centering
\phi_i=\frac{\gamma_1(n_1-4N)+\sqrt{\gamma^2n_1(8-7n_1)}}{4\gamma_1^2\kappa}\,
,
\end{equation}
Specifying the free parameters of the theory could produce results
compatible with the experimental values for the spectral indices
and the tensor-to-scalar ratio introduced in Eq. (\ref{results}).
Assuming that ($\omega$, $V_1$, $\Lambda_1$, $\lambda_1$,
$N$, $\gamma_1$, $n_1$)=(1, 1, 100, 1, 60, -100,
0.5), in reduced Planck units, so for $\kappa^2=1$, the model at
hand produces acceptable results, since $n_S=0.967045$ and
$r=0.000551$ are both compatible with the latest
observations\cite{Akrami:2018odb}. Moreover,the tensor spectral
index takes the value $n_T=0.000069$. Finally we mention that the
scalar field in equations (\ref{scalariA}) and (\ref{scalarfA})
takes the values $\phi_i=0.6025$ and $\phi_f=0.0025$ and clearly
shows that the field is decreasing as time flows. In Fig.
\ref{plot1}. we present the contour plots plot of two
observable quantities, namely $n_S$ and $r$, which is indicative
of the existence of more than a single set of values for the free
parameter which can lead to phenomenologically viable results.
Finally, we examine each approximation which was made in order to
derive the previous results. According to the previous set of
parameters in reduced Planck units always, during the first
horizon crossing, $\dot H\sim\mathcal{O}(10^{-5})$ and
$H^2\sim\mathcal{O}(10^{-3})$ so the slow-roll assumption holds
true. In addition,
$\frac{1}{2}\omega\dot\phi^2\sim\mathcal{O}(10^{-7})$ while
$V\sim\mathcal{O}(10^{-1})$ and lastly,
$\ddot\phi\sim\mathcal{O}(10^{-7})$ and
$H\dot\phi\sim\mathcal{O}(10^{-5})$. Hence, the slow-roll
conditions (\ref{slowrollapprox}) are valid. All that remains is
to ascertain the validity of the rest approximations. It turns out
that $\xi'H^4\sim\mathcal{O}(10^{-30})$ which is negligible
compared to $V'\sim\mathcal{O}(1)$ and therefore, the differential
equation of the scalar potential is justified. Furthermore,
$h'H^2\sim\mathcal{O}(10^{-3})$ so in principle, it could also be
omitted from the differential equation (\ref{Vdifeq1}). For
Hubble's parameter, $\dot\xi H^3\sim\mathcal{O}(10^{-32})$ and
$3H\dot h\sim\mathcal{O}(10^{-3})$, which are both much lower in
magnitude, compared to the scalar potential. Lastly, we verify the
last approximations made for the Hubble's rate derivatives. We
note that $\dot\xi H\dot H\sim\mathcal{O}(10^{-34})$, $H\dot
h\sim\mathcal{O}(10^{-3})$ and
$h''\dot\phi^2\sim\mathcal{O}(10^{-5})$ hence all the
approximations made are in fact valid.

Lastly we note that even if in Eq. (\ref{motion7}) we included
also the term $3H\dot h\kappa^{-2}$, meaning that if we were to
use the following equation
\begin{equation}
\centering
H^2\simeq\frac{\kappa^2V}{3h\left(1+\frac{h'}{h}\frac{\xi'}{\xi''}\right)}\,
,
\end{equation}
for the Hubble rate, we would end up with the exact same values
for the observed quantities. This property reassures us about the
chosen approach.

This is an interesting model due to the fact that in the minimal
case where $h(\phi)=1$ \cite{Odintsov:2020sqy}, it can be shown
that the model may lead to either eternal inflation or no
inflation at all. That result implies that the non-minimal case
provides a possible way of producing viable phenomenology, for a
coupling function which in the minimally coupled
Einstein-Gauss-Bonnet theory would lead to non-viable results.


\subsection{A Model with Power-Law $h(\phi)$ and Error $\xi(\phi )$ Functions}

Suppose now that the coupling functions are defined as,
\begin{equation}
\label{h2} \centering h(\phi)=\Lambda_2(\kappa\phi)^{n_2}\, ,
\end{equation}
\begin{equation}
\label{xi2} \centering
\xi(\phi)=\frac{2\lambda_2}{\sqrt{\pi}}\int_{0}^{\gamma_2\kappa\phi}{e^{-x^2}dx}\,
,
\end{equation}
where $x$ is an auxiliary integration variable. Similar to the
previous case, the coupling functions are chosen in a specific way
so that they lead to simplified ratios as shown below,
\begin{align}
\centering h'&=\frac{n_2}{\phi}h&\xi''&=-2\gamma_2^2\kappa^2\phi\xi'\,
.
\end{align}
Let us assume that Hubble's parameter and its derivative are
approximately equal to,
\begin{equation}
\label{motion9} \centering
H^2\simeq\frac{k^2V}{3\left(h+h'\frac{\xi'}{\xi''}\right)}\, ,
\end{equation}
\begin{equation}
\label{motion10} \centering
\dot{H}\simeq\frac{H^2}{2}\frac{h'}{h}\frac{\xi'}{\xi''}\, ,
\end{equation}
This is a different approximation in comparison to the one used in
the previous subsection, in which we kept more terms in Hubble's
parameter and less in its derivative. However we shall see that
for this particular model, either Eq. (\ref{motion7}) or Eq.
({\ref{motion9}) lead to viable results. Furthermore, due to the
approximations made in Hubble's derivative, the sign of slow-roll
index $\epsilon_1$ will be positive throughout this model.

The difference in the definition of Hubble's parameter in these
two models is that it leads to a much more complicated form of the
scalar potential. Assuming the previous definition, Eq.
(\ref{Vdifeq1}) produces the following form,
\begin{equation}
\label{potB} \centering V(\phi)=V_2(n_2-2(\gamma_2\kappa\phi)^2)^n
\exp\left(_2F_1\left(1, 1-\frac{n_2}{2}, 2-\frac{n_2}{2};
\frac{2(\gamma_2\kappa\phi)^2}{n_2}\right)\alpha_2(\kappa\phi)^{2-n_2}\right)\,
,
\end{equation}
where $\alpha_2=\frac{\omega}{\Lambda_2 n_2(n_2-2)}$, $V_2$ is a constant
with mass dimensions [m]$^{4}$ and $_2F_1\left(1, 1-\frac{n_2}{2},
2-\frac{n_2}{2}; \frac{2(\gamma_2\kappa\phi)^2}{n_2}\right)$ is the
hypergeometric function. This is a very complicated potential and
is a direct result of the extra term participating in Hubble's
parameter. As in the previous model, the slow-roll indices have
either extremely simple or very perplexed forms as shown below,
\begin{equation}
\label{index1B} \centering
\epsilon_1\simeq\frac{-n_2}{(2\gamma_2\kappa\phi)^2}\, ,
\end{equation}
\begin{equation}
\label{index2B} \centering
\epsilon_2\simeq\frac{2-n_2}{(2\gamma_2\kappa\phi)^2}\, ,
\end{equation}
\begin{equation}
\label{index3B} \centering
\epsilon_3\simeq\frac{-n_2}{(2\gamma_2\kappa\phi)^2}\, ,
\end{equation}
\begin{equation}
\label{index5B} \centering \epsilon_5\simeq\frac{16
(\gamma_2\kappa\phi) ^3 \lambda_2  \kappa ^4 V(\phi )-3 \sqrt{\pi }
\Lambda_2 ^2 n_2 e^{ (\gamma_2\kappa\phi) ^2} (\kappa  \phi )^{2 n_2}
\left(2 (\gamma_2\kappa\phi)^2-n_2\right)}{(2\gamma_2\kappa\phi)^2
\left(8 \gamma_2\kappa\phi \lambda_2  \kappa^4 V(\phi )+3
\sqrt{\pi } \Lambda_2 ^2 e^{(\gamma_2\kappa\phi)^2} (\kappa  \phi )^{2
n_2} \left(2 (\gamma_2\kappa\phi)^2-n_2\right)\right)}\, ,
\end{equation}
In this model, indices $\epsilon_4$ and $\epsilon_6$ are not
quoted, due to their complicated form. In contrast to the previous
model, now indices $\epsilon_1$ and $\epsilon_3$ are equivalent
and moreover, the form of the first three slow-roll indices
appears to be a very simple equation, especially compared to index
$\epsilon_5$ which is also depending on the scalar potential. In
addition, the form of $\epsilon_1$ greatly constrains the values
available for the exponent $n$ of the coupling scalar function
$h(\phi)$ since it can take only negative values.

Similar to the previous, model, the value of the scalar field at
the end of inflation is derived from the equation $\epsilon_1=1$
and therefore it reads,
\begin{equation}
\label{scalarfB} \centering
\phi_f=\pm\frac{\sqrt{-n_2}}{2\gamma_2\kappa}\, ,
\end{equation}
As a result, the equation of the $e$-foldings number generates the
formula of the value of the scalar field at the initial stage of
inflation, which in tern is written as,
\begin{equation}
\label{scalariB} \centering
\phi_i=\pm\frac{\sqrt{4N-n_2}}{2\gamma_2\kappa}\, ,
\end{equation}
Assuming that in reduced Planck Units, the free parameters have
the following values ($\omega$, $V_2$, $\Lambda_2$,
$\lambda_2$, $N$, $\gamma_2$, $n_2$)=(1, -1, -4, 1.5$\cdot10^{31}$,
60, -10, -3), we obtain viable results for the observational
quantities, which are in good agreement with experimental evidence
\cite{Akrami:2018odb}, since $n_s=0.964397$ and
$r=8.3836\cdot10^{-6}$. In addition, the tensor spectral index
generates the value $n_T=-0.779423$. However, the compatibility
with the observational data for the model at hand can come for a
wide range of the free parameters values, as can also be seen in
Figs. \ref{plot2}.
\begin{figure}[h!]
\centering
\includegraphics[width=17pc]{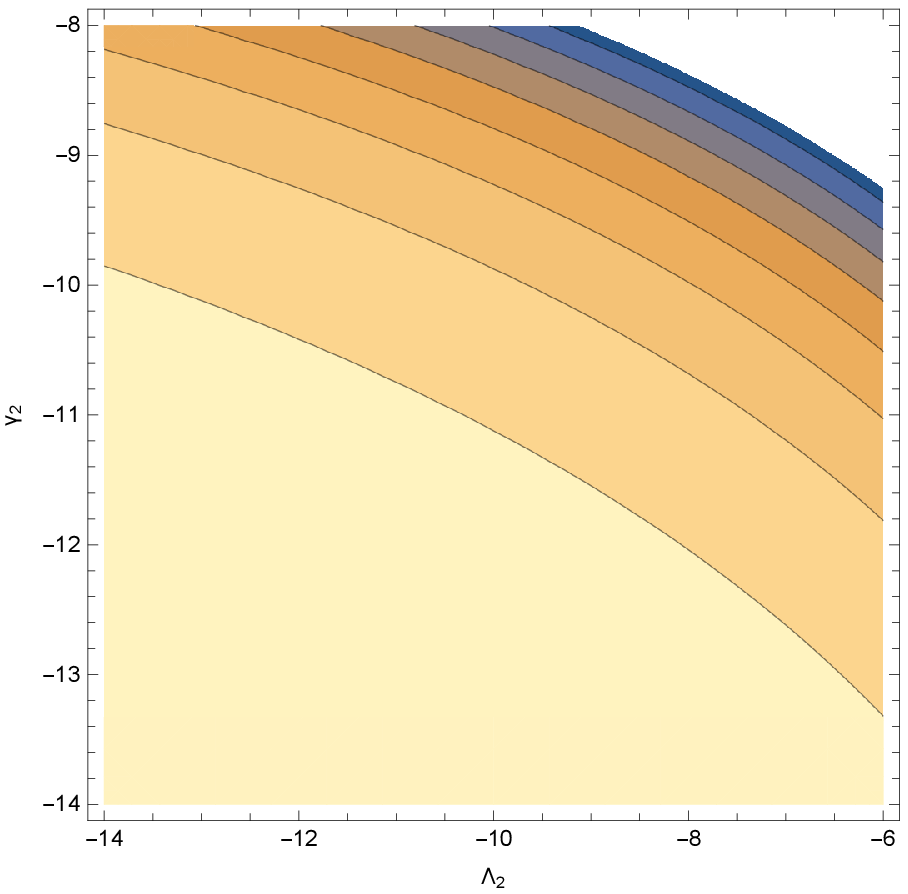}
\includegraphics[width=3pc]{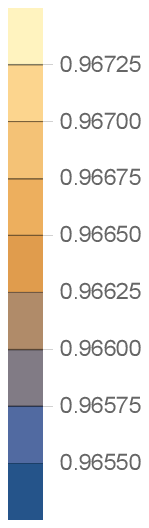}
\includegraphics[width=17pc]{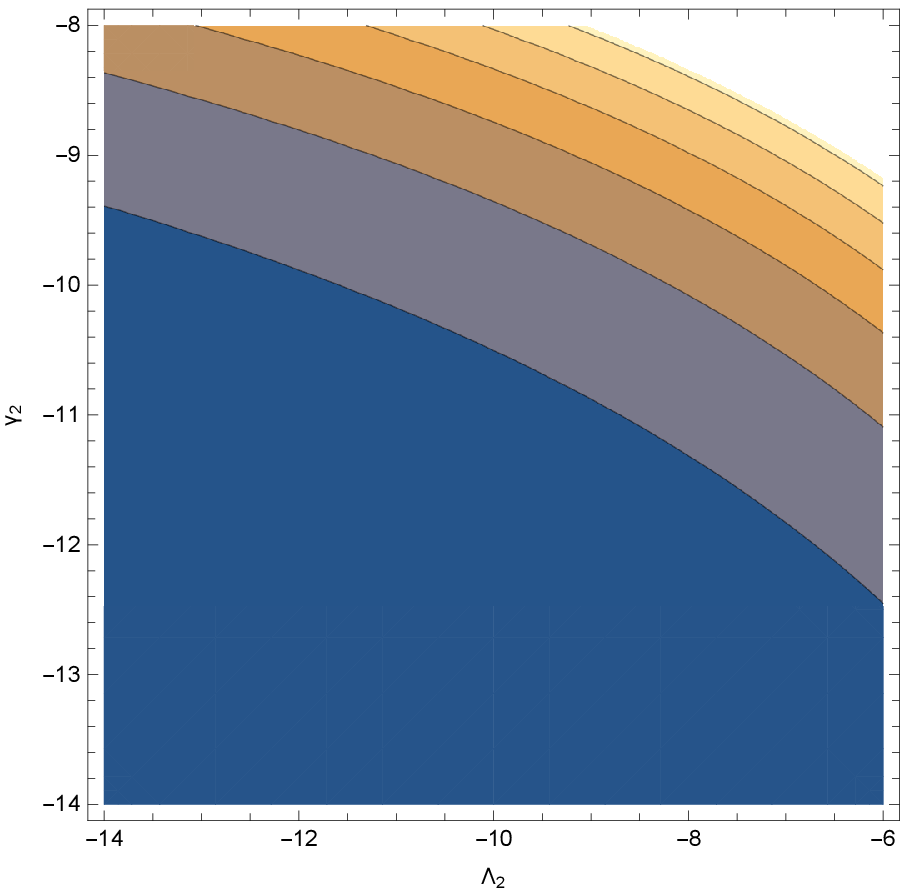}
\includegraphics[width=3pc]{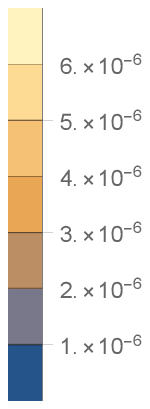}
\caption{Contour plot of spectral index $n_S$ (left) and
tensor-to-scalar ratio $r$ (right) depending on parameters
$\Lambda_2$ and $\gamma_2$, ranging from [-14, -6] and [-14,-8]
respectively. Exponent $n$, although it affects greatly these
observed quantities, was left fixed at the value -3 simply due to
the hypergeometric function in the scalar potential, which would
produce complex number if $n$ was not a negative integer.}
\label{plot2}
\end{figure}
The fascinating result about this model is that with an
appropriate fine-tuning, the value of the tensor-to scalar ratio
can drop drastically. For instance, changing only the values for
parameters $\Lambda_2$ and $\lambda_2$ to $\Lambda_2=-4\cdot10^{10}$ and
$\lambda_2=1.5\cdot10^{13}$ leads to the values $n_S=0.96748$ and
$r=8.38195\cdot10^{-44}$.  This implies that the model can survive
many, if not any restriction on the tensor-to-scalar ratio
generated by better experiments conducted in the future, although
it is rather impossible to detect any tensor modes for such a
small $r$.

Finally, we examine the validity of the approximations made for
this model, during the era of inflation. At the initial moment of
inflation, we note that $\dot H\sim\mathcal{O}(10^{-8})$ and
$H^2\sim\mathcal{O}(10^{-10})$,
$\frac{1}{2}\omega\dot\phi^2\sim\mathcal{O}(10^{-13})$ and
$V\sim\mathcal{O}(10^{-7})$ and finally,
$\ddot\phi\sim\mathcal{O}(10^{-12})$, in reduced Planck units,
while $H\dot\phi\sim\mathcal{O}(10^{-10})$, hence the slow-roll
approximations (\ref{slowrollapprox}) are still valid even if the
magnitudes are not separated by many orders. Concerning the
approximations in the differential equation of the scalar
potential and Hubble's parameter, we mention that
$12\xi'H^4\sim\mathcal{O}(10^{-9})$ and $12\dot\xi
H^3\sim\mathcal{O}(10^{-11})$ , so compared to
$V'\sim\mathcal{O}(10^{-6})$ and $V\sim\mathcal{O}(10^{-7})$
respectively, our approximations are once again justified.
Finally, we note that $\dot\phi^2$ and $8\dot\xi H\dot H$ are
both or order $\mathcal{O}(10^{-13})$
,$h''\dot\phi^2\sim\mathcal{O}(10^{-10})$ and $H\dot
h\sim\mathcal{O}(10^{-8})$ hence all the approximations which were
made for this model apply.

As a last comment, it is worth mentioning the selection of
Hubble's parameter in Eq, (\ref{motion9}). Genuinely, speaking, it
was chosen not because it is necessary in order to achieve
viability, but in order to deviate from the approach of the
previous model, although if one were to use equations
(\ref{motion7}) and (\ref{motion10}) could also generate results
compatible with the observations, with a slightly different
specification of the free parameters of the theory. Many relations
however shall remain the same. For instance, we mention that
slow-roll indices $\epsilon_1$ through $\epsilon_3$ remain
invariant since they are depending solely on the ratios of the
coupling functions. This is also the reason why we choose such
coupling functions in the first place. On the other hand, the
indices $\epsilon_5$ and $\epsilon_6$ certainly change, not just
due to the difference in the Hubble rate, but also due to the
scalar potential, since the latter is derived directly from the
first. Instead of explaining it, it is better to show it using
equations. Using equations (\ref{motion7}) and (\ref{motion10}),
meaning that,
\begin{equation}
\centering H^2\simeq\frac{\kappa^2 V}{3h}\, ,
\end{equation}
\begin{equation}
\centering \dot H=\frac{H^2}{2}\frac{h'}{h}\frac{\xi'}{\xi''}\, ,
\end{equation}
the scalar potential which is derived from Eq. (\ref{Vdifeq1}) is,
\begin{equation}
\centering
V(\phi)=V_2(\kappa\phi)^{2n_2}\exp\left(-\frac{\omega(\kappa\phi)^{-n_2}}{2n_2\gamma_2^2\Lambda_2}\right)\,
.
\end{equation}
This is obviously a much simpler form compared to Eq. (\ref{potB})
which can be attributed to the simpler expression of the Hubble
rate. Consequently,
\begin{equation}
\centering \epsilon_5\simeq\frac{8 \gamma_2\kappa\phi \lambda_2
\kappa^4  V(\phi )-3 \sqrt{\pi } \Lambda_2 ^2 n_2
e^{(\gamma_2\kappa\phi)^2} (\kappa  \phi )^{2 n_2}}{16 \gamma_2
\kappa\phi \lambda_2  \kappa^4 V(\phi )+12 \sqrt{\pi }
(\Lambda_2\gamma_2\kappa\phi)^2 e^{(\gamma_2\kappa \phi)^2} (\kappa
\phi )^{2 n_2}}\, ,
\end{equation}
\begin{equation}
\centering \epsilon_6\simeq\frac{4 \lambda_2 \kappa^4 V(\phi )
\left(2 (\gamma_2\kappa \phi)^2+n_2+1\right)-\kappa\phi  \left(4
\lambda_2 \kappa^3 V'(\phi )+3 \sqrt{\pi } \gamma_2  \Lambda_2 ^2 n_2
e^{(\gamma_2 \kappa \phi)^2} (\kappa  \phi )^{2 n_2}\right)}{(2
\gamma_2\kappa\phi)^2 \left(4 \lambda_2  \kappa^4V(\phi )+3
\sqrt{\pi } \Lambda_2 ^2\gamma_2 \kappa\phi  e^{(\gamma_2\kappa\phi)^2}
(\kappa  \phi )^{2 n_2}\right)}\, ,
\end{equation}
The change of the Hubble rate, instead of altering the potential
and the last two slow-roll indices, will lead to a different set
of free parameters which lead to compatibility. For instance,
assuming that the only change is $\lambda_2=1.5\cdot10^{21}$, the
spectral indices and the tensor-to-scalar ratio (\ref{results})
take the values $n_S=0.967698$, $n_T=3.72809\cdot10^{-8}$ and
$r=7.39423\cdot10^{-9}$ which are accepted values
\cite{Akrami:2018odb} as well. Lastly, the approximations made
still hold true but instead of maintaining the same order of
magnitude, the keep their relative order. For instance,
$X/V\sim\mathcal{O}(10^{-6})$ and $\dot
H/H^2\sim\mathcal{O}(10^{-2})$, exactly as in the previous
approach.

Hence, both approaches may lead to viable results. Neglecting the
second term in the denominator of Eq. (\ref{motion9}) leads only
to a change of a single parameter and therefore, the numerical
value of quantities such as the Hubble parameter itself. However a
single redefinition of a parameter is capable of restoring the
viability.

\subsection{A Model with Trigonometric $h(\phi)$ and Power-law $\xi(\phi )$ Functions}

Let us now present an apparently elegant model, however non-viable
since the slow-roll approximation breaks down. In this particular
model, we shall assume basic functions as coupling functions,
which at first sight might seem odd. Let,
\begin{equation}
\label{h3} \centering h(\phi)=\Lambda_3
\sin(\gamma_3\kappa\phi+\theta)\, ,
\end{equation}
\begin{equation}
\label{xi3} \centering \xi(\phi)=\lambda_3(\kappa\phi)^{n_3}\, .
\end{equation}
This choice benefits us due to the connection between the
derivatives of such functions as shown below,
\begin{align}
\centering h''&=-(\gamma_3\kappa)^2h&\xi''&=\frac{n_3-1}{\phi}\xi'\, ,
\end{align}
In this model, we shall assume that Hubble's parameter and its
derivative are given by the following expressions,
\begin{equation}
\label{motion11} \centering
H^2\simeq\frac{k^2V}{3h'\frac{\xi'}{\xi''}}\, ,
\end{equation}
\begin{equation}
\label{motion12} \centering \dot
H\simeq\frac{-H^2}{2}\frac{h''}{h}\left(\frac{\xi'}{\xi''}\right)^2\,
.
\end{equation}
Despite the odd choice for the coupling functions, it is obvious
that it facilitates our study since the ratios are simplified and
as a result, slow-roll index $\epsilon_1$ shall be simplified as
well. Furthermore, we shall consider this particular index has a
positive value, i.e $\epsilon_1=\frac{\dot H}{H^2}$. Before
continuing to the evaluation of the slow-roll indices, let us
first derive the formula for the scalar potential from Eq.
(\ref{Vdifeq1}). According to the previous designations, the
scalar potential must have the following form,
\begin{equation}
\label{pot3} \centering
V(\phi)=V_3(\kappa\phi)^{2(n_3-1)}\left(\cos\left(\frac{\gamma_3\kappa\phi+\theta}{2}\right)-\sin\left(\frac{\gamma_3\kappa\phi+\theta}{2}\right)\right)^{\alpha_3}\left(\cos\left(\frac{\gamma_3\kappa\phi+\theta}{2}\right)+\sin\left(\frac{\gamma_3\kappa\phi+\theta}{2}\right)\right)^{-\alpha_3}\,
,
\end{equation}
where $\alpha_3=\frac{\omega}{\Lambda_3\gamma_3^2}$. As shown, the
scalar potential has again a very complicated form and cannot be
used easily. Despite the form, the scalar potential participates
only in slow-roll indices $\epsilon_4$ through $\epsilon_6$ and
only the first three indices should concern us. More specific, the
slow-roll indices can be written as,
\begin{equation}
\label{index1C} \centering
\epsilon_1\simeq\frac{1}{2}\left(\frac{\gamma_3\kappa\phi}{n_3-1}\right)^2\,
,
\end{equation}
\begin{equation}
\label{index2C} \centering
\epsilon_2\simeq\frac{1}{n_3-1}+\frac{1}{2}\left(\frac{\gamma_3\kappa\phi}{n_3-1}\right)^2\,
,
\end{equation}
\begin{equation}
\label{index3C} \centering \epsilon_3\simeq\frac{\gamma_3\kappa\phi
Cot(\gamma_3\kappa\phi+\theta)}{2(n_3-1)}\, ,
\end{equation}
\begin{equation}
\label{index5C} \centering \epsilon_5\simeq\frac{4 \kappa ^3
\lambda_3  (n_3-1) n_3 V(\phi ) (\kappa  \phi )^{n_3}-3 \gamma_3 ^2 \kappa
\Lambda_3 ^2 \phi ^2 \cos ^2(\gamma_3  \kappa  \phi +\theta )}{(n_3-1)
\left(8 \kappa ^3 \lambda_3  n_3 V(\phi ) (\kappa  \phi )^{n_3}-3
\gamma_3  \Lambda_3 ^2 \phi  \sin (2 (\gamma_3  \kappa  \phi +\theta
))\right)}\, ,
\end{equation}
\begin{equation}
\label{index6C} \centering \epsilon_6\simeq\frac{\kappa  \phi
\left(4 \lambda_3  n_3 (\kappa  \phi )^{n_3} \kappa^3V'(\phi )-3
\gamma_3 ^2 \Lambda_3 ^2 \kappa\phi  \cos ^2(\gamma_3  \kappa  \phi
+\theta )\right)+4 \lambda_3  n_3\kappa^4 V(\phi ) (\kappa  \phi
)^{n_3} (\gamma_3  \kappa  \phi  \tan (\gamma_3  \kappa  \phi +\theta
)+n_3-1)}{(n_3-1) \left(8 \lambda_3  n_3\kappa^4 V(\phi ) (\kappa
\phi )^{n_3}-3 \gamma_3  \Lambda_3 ^2 \kappa\phi  \sin (2 (\gamma_3  \kappa
\phi +\theta ))\right)}\, ,
\end{equation}
Similarly, index $\epsilon_4$ was omitted due to its complicated
form. In this model, we see that index $\epsilon_3$ has a
$\phi$-dependence. Apparently, in this approach the slow-roll
indices $\epsilon_3$ through $\epsilon_6$ keep oscillating during
the era of inflation. The frequency of the oscillations is
obviously depending on parameter $\gamma$.

According to our previous statements, the final value of the
scalar field has a particularly simple form due to simple
expression of slow-roll index $\epsilon_1$, and similarly the
initial value presumably is described in simple terms. Both values
are shown below respectively,
\begin{equation}
\label{scalarfC} \centering
\phi_f=\pm\frac{\sqrt{2}}{\kappa}\left|\frac{n_3-1}{\gamma_3}\right|\, ,
\end{equation}
\begin{equation}
\label{scalariC} \centering
\phi_i=\pm\frac{\sqrt{2}}{\kappa}\left|\frac{n_3-1}{\gamma_3}\right|e^{-\frac{N}{n_3-1}}\, ,
\end{equation}
In the following, we shall choose positive signs for both values.

Assuming that in reduced Planck Units, ($\omega$, $V_3$,
$\Lambda_3$, $\lambda_3$, $N$, $\theta$, $\gamma_3$, $n_3$)=(1,
$10^{-6}$, $10^{3}$, $10^{13}$, 60, $\frac{\pi}{3}$, 1, 14.5) the
values for the observed quantities derived from Eq.
(\ref{results}) are in agreement with observations
\cite{Akrami:2018odb}, as $n_S=0.96405$ and $r=0.0388044$ are both
compatible values. Moreover, the tensor spectral index takes the
value $n_T=-0.00485$ and the scalar field takes the values
$\phi_i=0.224208$ and $\phi_f=19.0919$. In this case, the field
increases with time.

Despite the elegance or the accuracy of this model, it is in fact
a non-viable model due to the approximations imposed. For
simplicity, we shall not mention the order of magnitude of each
object, but we shall state that even though the slow-roll
approximations (\ref{slowrollapprox}) do apply, the term $H\dot h$
is greater than $h''\dot\phi^2$ and therefore, our approach with
Eq. (\ref{motion12}) is rendered invalid. Numerically speaking,
$H\dot h\sim\mathcal{O}(10^{-25})$ while
$h''\dot\phi^2\sim\mathcal{O}(10^{-26})$. Perhaps a different set
of parameters or a complete different model could under the same
assumptions yield a viable description for the inflationary era.
Such possibility was not further studied.

\subsection{A Model with Linear $h(\phi)$ and Exponential $\xi(\phi )$ Functions}

As a final model, we shall make the almost the same choices for
the functions $h(\phi)$ and $\xi(\phi )$ as in the first model we
presented, but in this case, we shall assume a linear form for
$h(\phi)$. The reason is that this specific linear form of
$h(\phi)$ has a direct effect on the tensor-to-scalar ratio as we
show shortly. The coupling functions shall take the form,
\begin{equation}
\label{h5} \centering h(\phi)=\Lambda_4\kappa\phi\, ,
\end{equation}
\begin{equation}
\label{xi5} \centering \xi(\phi)=\lambda_4 e^{\gamma_4\kappa\phi}\, .
\end{equation}
As it was demonstrated in a previous subsection, this choice leads
to simplified ratios of the derivatives of the coupling functions.
This time however, instead of a power-law model, we chose a linear
model simply because the second derivative of the coupling
function $h(\phi)$ is set equal to zero. Let us assume that the
Hubble rate and its derivatives are approximated as follows,
\begin{equation}
\label{motion15} \centering
H^2\simeq\frac{k^2V}{3h\left(1+\frac{h'}{h}\frac{\xi'}{\xi''}\right)}\,
,
\end{equation}
\begin{equation}
\label{motion16} \centering \dot
H\simeq\frac{H^2}{2}\frac{\xi'}{\xi''}\left(\frac{h'}{h}-\frac{\omega\kappa^2}{h}\frac{\xi'}{\xi''}\right)\,
.
\end{equation}
Due to the form of $\dot H$, it is easier to assume the positive
value of index $\epsilon_1$, meaning that $\epsilon_1=\frac{\dot
H}{H^2}$. Furthermore, this approach in fact could be categorized
as a special case of Eq. (\ref{motion5}) where we neglect only the
term $8c_1\dot\xi H\dot H$ as the term proportional to $h''$
disappears due to the linear choice for the coupling function. Let
us proceed with the evaluation of the scalar potential. According
to equation (\ref{Vdifeq1}), the potential must be equal to,
\begin{equation}
\label{potE} \centering
V(\phi)=V_4\Lambda_4(1+\gamma_4\kappa\phi)^{2-\alpha_4}\, ,
\end{equation}
where $V_4$ is the integration constant with mass dimensions
[m]$^{4}$ and $\alpha_4=\frac{\omega}{\gamma_4\Lambda_4}$. This is by
far the simplest scalar potential that was derived in this paper,
due to the choice of the coupling scalar functions. In the
following, we present the slow-roll indices which are expected to
have very simplified forms, compared to the previous models, and
indeed these are,
\begin{equation}
\label{index1E} \centering
\epsilon_1\simeq\frac{1}{2\gamma_4}(1-\frac{\omega}{\gamma_4\Lambda_4})\frac{1}{\kappa\phi}\,
,
\end{equation}
\begin{equation}
\label{index2E} \centering \epsilon_2\simeq\frac{\gamma_4  \Lambda_4
-\omega }{2 \gamma_4 ^2 \kappa  \Lambda_4  \phi }\, ,
\end{equation}
\begin{equation}
\label{index3E} \centering
\epsilon_3\simeq\frac{1}{2\gamma_4\kappa\phi}\, ,
\end{equation}
\begin{equation}
\label{index5E} \centering \epsilon_5\simeq\frac{4 \gamma_4 ^2 
\lambda_4  \kappa^4V(\phi ) e^{\gamma_4  \kappa  \phi }-3 \Lambda_4 ^2
(\gamma_4  \kappa  \phi +1)}{2 \gamma_4    \left(4 \gamma_4 \lambda_4
\kappa^4 V(\phi ) e^{\gamma_4  \kappa  \phi }-3 \Lambda_4 ^2
\kappa\phi  (\gamma_4  \kappa \phi +1)\right)}\, ,
\end{equation}
\begin{equation}
\label{index6E} \centering \epsilon_6\simeq\frac{(\gamma_4  \kappa
\phi +1) \left(4 \gamma_4  \lambda_4  e^{\gamma_4  \kappa  \phi }
\kappa^3V'(\phi )-3 \Lambda_4 ^2 (\gamma_4  \kappa  \phi +1)\right)+4
\gamma_4 ^3 \kappa\phi \lambda_4  \kappa^4  V(\phi ) e^{\gamma_4
\kappa  \phi }}{2 \gamma_4   (\gamma_4  \kappa  \phi +1) \left(4
\gamma_4 \lambda_4  \kappa^4V(\phi ) e^{\gamma_4  \kappa  \phi
}-3 \Lambda_4 ^2\kappa \phi  (\gamma_4  \kappa  \phi +1)\right)}\, ,
\end{equation}
Even in the simple linear form of the scalar coupling function
$h(\phi)$, the index $\epsilon_4$ has a quite lengthy form so we
did not quote it here. However, the first three indices are simple
as expected and thus the values of the scalar field can be easily
derived. Thus, similar to previous models,
\begin{equation}
\label{scarafE} \centering \phi_f=\frac{\gamma_4  \Lambda_4 -\omega
}{2 \gamma_4 ^2 \kappa  \Lambda_4 }\, ,
\end{equation}
\begin{equation}
\label{scalariE} \centering \phi_i=\frac{\gamma_4 ^2 \Lambda_4 -2
\gamma_4  \Lambda_4  N-\omega }{2 \gamma_4 ^2 \kappa  \Lambda_4 }\, ,
\end{equation}
Assuming that the free parameters of the model obtain the values
($\omega$, $V_4$, $\Lambda_4$, $\lambda_4$, $N$, $\gamma_4$)=(
1, 1.4, -40, 1, 60, 10) in reduced Planck Units, meaning
$\kappa=1$, then it turns out that Eq. (\ref{results}) produces
compatible results \cite{Akrami:2018odb} since $n_S=0.964796$ and
$r=0.000363645$. Furthermore the tensor spectral index is equal to
$n_T=-0.00004587$. In Fig. \ref{plot4} we present the parametric
plot of the $n_S$ and $r$. This case is quite different from the
first model since here there exists only an one-on-one correlation
between $n_S$ and $r$.
\begin{figure}[h!]
\centering
\includegraphics[width=20pc]{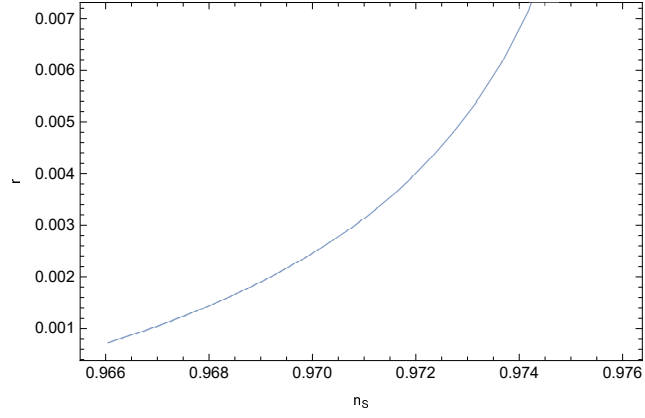}
\caption{Parametric plot of spectral index of scalar perturbations
$n_s$ (x axis) and tensor-to-scalar ratio $r$ (y axis) depending
on parameters $\Lambda_4$ and $\gamma_4$ ranging from [-10,-1] and
[1,20] respectively. It shows a clearly a one-on-one connection
between the observed quantities which is attributed to the model
of the coupling functions $h(\phi)$ and $\xi(\phi)$ and not the
approach on Hubble's parameters.} \label{plot4}
\end{figure}

Finally, we examine the validity of our approximations. For the
slow-roll approximations (\ref{slowrollapprox}), we note that
$\dot H\sim\mathcal{O}(10)$ and $H^2\sim\mathcal{O}(10^4)$, the
kinetic term is of order
$\frac{1}{2}\omega\dot\phi^2\sim\mathcal{O}(10)$ while the scalar
potential is of order $V\sim\mathcal{O}(10^6)$ and finally
$\ddot\phi\sim\mathcal{O}(1)$ while
$H\dot\phi\sim\mathcal{O}(10^3)$ so clearly the slow-roll
conditions do apply, even marginally. The rest approximations
concern this specific approach and in fact do apply since for the
differential equation of the scalar potential,
$V'\sim\mathcal{O}(10^{6})$ while
$12H^4\sim\mathcal{O}(10^{-15})$, for Hubble's parameter, we
note that $V\sim\mathcal{O}(10^6)$ while $12\dot\xi
H^3\sim\mathcal{O}(10^{-15})$ and finally for Hubble's derivative,
we mention that $8\dot\xi H\dot H\sim\mathcal{O}(10^{-17})$
while $\dot\phi^2\sim\mathcal{O}(10^{2})$. The orders of magnitude
indicate that each approximation made in this approach is in fact
valid. In addition, this model contained the least amount of extra
approximations necessary to make the system of equations of motion
solvable due to the linear choice of the function coupled to the
Ricci scalar, as only a single extra approximation, apart from the
slow-roll approximations, was made in each equation of motion.

\section{Overview of Generalized Slow-roll Conditions for GW170817-compatible non-minimally Coupled Einstein-Gauss-Bonnet Gravity}

In the final section of this paper, we shall present all the
possible approximations that can be made for an
GW170817-compatible non-minimally coupled Einstein-Gauss-Bonnet
gravity, in order to obtain a viable inflationary era. Apart from
the ones used in the models presented in the previous section,
there exist also other possible arrangements of approximations
which we shall present without supporting them with a model but
state the necessary conditions under which one could work.
Firstly, the third equation of motion, Eq. (\ref{motion6}) in this
framework is more or less acting like an auxiliary equation which
generates the scalar potential once the coupling functions and the
form of Hubble's parameter have been specified. Therefore, one may
wish to keep that equation as it is, without altering it.
Obviously this is acceptable, but we made the approximation
$12\xi'H^4\ll V'$ in order to simplify the expression. In
consequence, there exist two possibilities. One can either work
with the full expression,
\begin{equation}
\centering
V'+3H^2\left(\omega\frac{\xi'}{\xi''}-2\frac{h'}{\kappa^2}+4\xi'H^2\right)=0\,
,
\end{equation}
or the simplified expressions,
\begin{equation}
\centering
V'+3H^2\left(\omega\frac{\xi'}{\xi''}-2\frac{h'}{\kappa^2}\right)=0\,
,
\end{equation}
\begin{equation}
\centering
\label{V'2}
V'-6H^2\frac{h'}{\kappa^2}=0\,
,
\end{equation}
Only the first form was used in this letter. For the sake of consistency, 
we mention that the second model is in fact viable in both cases.
This approximation simplifies the
differential equation greatly and the scalar potential can be
extracted easier. Furthermore, as it was demonstrated in the
previous models, it is a decent assumption since compared to the
rest terms, $12\xi'H^4$ is in fact negligible.

Concerning Hubble's Parameter in Eq. (\ref{motion4}), the relation
can be simplified in three ways,
\begin{equation}
\label{H1} \centering H^2\simeq\frac{\kappa^2V}{3h}\, ,
\end{equation}
\begin{equation}
\centering \label{H3} H^2\simeq\frac{\kappa^2
V}{3h'}\frac{\xi''}{\xi'}\, ,
\end{equation}
\begin{equation}
\label{H2} \centering
H^2\simeq\frac{\kappa^2V}{3h\left(1+\frac{h'}{h}\frac{\xi'}{\xi''}\right)}\,
,
\end{equation}
The first approach is similar to the second one but is also
simplified and therefore leads to a better solution for the scalar
potential. There is no point in adding the kinetic term since it
is many orders smaller, than the potential in the slow-roll
approximation. Furthermore, the term $\xi'\dot\phi H^4$ is also
negligible compared to the other terms. The dominant contribution
comes of course from the scalar potential and the squared Hubble
rate. For the sake of consistency however, we mention that since
the slow-roll conditions are assumed to hold true, then $\epsilon_3\ll1$
and as a result Eq. (\ref{H3}) is invalid in this regime. Indeed, in the 
trigonometric choice it was shown that even though compatible with the 
observations results where produced, the model was intrinsically unrealistic.
This is an expected feature under the slow-roll assumptions.

On the other hand, since Hubble's derivative is written
proportionally to Hubble's parameter squared, the slow-roll index
$\epsilon_1$ can be extracted directly from the form of Eq.
(\ref{motion5}), so the inflationary phenomenology strongly
depends on the approximations made in this particular equation.
The possible approximations are shown below,
\begin{equation}
\label{H'1} \centering \dot
H\simeq\frac{H^2}{2}\frac{h'}{h}\frac{\xi'}{\xi''}\, ,
\end{equation}
\begin{equation}
\label{H'2} \centering \dot
H\simeq-\frac{H^2}{2}\frac{h''}{h}\left(\frac{\xi'}{\xi''}\right)^2\,
,
\end{equation}
\begin{equation}
\label{H'3} \centering \dot
H\simeq-H^2\frac{\kappa^2\omega}{2h}\left(\frac{\xi'}{\xi''}\right)^2\,
,
\end{equation}
\begin{equation}
\label{H'4} \centering \dot
H\simeq\frac{H^2}{2}\frac{\xi'}{\xi''}\left(\frac{h'}{h}-\frac{h''}{h}\frac{\xi'}{\xi''}\right)\,
,
\end{equation}
\begin{equation}
\label{H'5} \centering \dot
H\simeq\frac{H^2}{2}\frac{\xi'}{\xi''}\left(\frac{h'}{h}-\frac{\kappa^2\omega}{h}\frac{\xi'}{\xi''}\right)\,
,
\end{equation}
\begin{equation}
\label{H'6} \centering \dot
H\simeq-\frac{H^2}{2}\left(\frac{\kappa^2\omega}{h}+\frac{h''}{h}\right)\left(\frac{\xi'}{\xi''}\right)^2\,
.
\end{equation}
Adding more terms than these in the equations above, renders the
system extremely difficult to solve, if not unsolvable. The term
$8\dot\xi H\dot H$ was neglected in most approximations, along
with the term $\ddot\phi h'$. The latter is easily justified from
the slow-roll approximation, since $\ddot\phi\ll H\dot\phi$.
Consequently, the term $\ddot\phi h'$ may be neglected in every
approximation, whether the term $Hh'\dot\phi$ participates in the
equation of motion or is also neglected. However they could be
used together, for instance if the constant-roll condition is
used. This case however would also change the form of $\dot\phi$
derived from the constraint of the velocity of gravitational
waves, so this is a topic of another study. The first is neglected
since it does not contain the coupling function $h(\phi )$, so it
will lead to difficulties. The same argument may be used for the
kinetic term which participates in the last two cases. However,
these cases could lead to a simple expression, perhaps a
polynomial, as was the case with the first model of the previous
section, so it was worth mentioning them as well. In each case,
the expression of Hubble's derivative is greatly simplified if the
coupling functions are chosen in an elegant way so that the ratios
of their derivatives are functionally simple. That was the reason
behind choosing simple functions such as power-law and exponential
coupling functions. Furthermore, it can easily be inferred that equations
(\ref{H'2}) and (\ref{H'6}) are incompatible with the slow-roll conditions
therefore they cannot be implemented in the present framework. Therefore
the overall approximated forms of $\dot H$ are 4. We shall return to this 
statement in the following.

The choice of Hubble's parameter does not alter the first three
slow-roll indices, but only affects the scalar potential derived
from Eq. (\ref{Vdifeq1}) and the value of the free parameters of
the theory which lead to phenomenologically viable results, as it
was demonstrated in the second model of the previous section.
Hence, it would be legitimate to work with a more inclusive
equation, such as Eq. (\ref{H2}) even if the second term of the
denominator is negligible, so long it leads to a manageable scalar
potential and not to physical inconsistencies. For instance we
mention that in the third model presented in the previous section,
the choice of a more inclusive Hubble parameter led to the
appearance of complex number, both in the observed quantities but
also in other quantities, such as the scalar potential and
consequently the Hubble rate itself. Moreover, these complex
numbers could not disappear with the choice of a better selection
of parameters or a different fine tuning. However, when switched
to Eq. (\ref{H1}), the complex numbers disappeared. Despite not
being a viable model at all, the third model of the previous
section is indicative of how intricate the system of the
gravitational equations of motion really is.

Let us continue with our study and refer to the conditions under
which the model is rendered viable in terms of the approach which
was chosen from the previous possible cases. Firstly, each
equation in this paper was derived by assuming the slow-roll
conditions hold true. This in turn implies that no matter the
choice of coupling functions or the extra approximations on the
equations of motion, one must ascertain whether the following
conditions are true during the first horizon crossing, or in other
words the initial moment of inflation.
\begin{align}
\centering \dot H &\ll H^2& \frac{1}{2}\omega\dot\phi^2&\ll
V&\ddot\phi&\ll H\dot\phi\, .
\end{align}
These approximations are essential and must hold true, otherwise
the whole approach would be rendered invalid. In addition to these
approximations, one must check whether the approximation made in
order to derive the differential equation (\ref{Vdifeq1}) and
(\ref{H1}) apply as well. These approximations are,
\begin{align}
\centering 12\xi'H^4&\ll V' & 12\dot\xi H^3&\ll V & 3H\dot
h&\ll\kappa^2V&h&\ll h'\, &\frac{\xi'}{\xi''}&\ll\frac{h'}{\kappa^2} .
\end{align}
Note that the third relation may be violated only if Eq.
(\ref{H2}) is used whereas the fourth refers only to the approach
of Eq. (\ref{H3}) which as mentioned previously is at variance with the slow-roll conditions
. Also, the last refers only to Eq. (\ref{V'2}). Finally, 
we mention the approximations which
must be valid in the equation of Hubble's derivative. For equation
(\ref{H'1}),
\begin{align}
\centering \kappa^2\omega\dot\phi^2&\ll H\dot h&h''\dot\phi^2&\ll
H\dot h& 8\kappa^2\dot\xi H\dot H&\ll H\dot h\, .
\end{align}
Concerning Eq. (\ref{H'2}),
\begin{align}
\centering \kappa^2&\ll h''& H\dot h&\ll h''\dot\phi^2 &
8\kappa^2\dot\xi H\dot H&\ll h''\dot\phi^2\, .
\end{align}
Similarly, using Eq. (\ref{H'3}) leads to the following
approximations,
\begin{align}
\centering h''&\ll\kappa^2& H\dot h&\ll \kappa^2\omega\dot\phi^2&
8\dot\xi H\dot H&\ll \omega\dot\phi^2\, ,
\end{align}
The rest approaches contain more terms and therefore lesser
approximations must be implemented. For instance, in order to use
Eq. (\ref{H'4}), the following approximations must be valid,
\begin{align}
\centering 8\kappa^2\dot\xi H\dot
H+\kappa^2\omega\dot\phi^2&\ll H\dot h+h''\dot\phi^2\, ,
\end{align}
and similarly, for Eq. (\ref{H'5}),
\begin{align}
\centering 8\kappa^2\dot\xi H\dot H+h''\dot\phi^2&\ll H\dot
h+\kappa^2\omega\dot\phi^2\, .
\end{align}
Finally, the necessary extra condition under which Eq. (\ref{H'6})
would be valid is,
\begin{align}
\centering H\dot h+8\kappa^2\dot\xi H\dot H&\ll
(\kappa^2\omega+h'')\dot\phi^2\, .
\end{align}
The last three approximations were written in this form for
convenience, but we note that a single expression on the right
hand side must be compared to each term of the left hand side
separately. Furthermore, each approximation made for either model
refers to the absolute value of each term. As mentioned before, the 
approximated forms of equations (\ref{H'2}) and (\ref{H'6}) cannot be 
used under the slow-roll assumption since they violate the expression
$\ddot h\ll H\dot h$.

No matter the choice of equations, if a single approximation is
invalid then the whole model will be wrong, even if the results
happen to be compatible with the latest observations, as was the
case with the third model the previous section.

\section{Conclusions}

In this paper, we presented a new approach on non-minimally
coupled theories of gravity which contain string corrections, by
imposing the condition that the gravitational wave speed is equal
to unity. We demonstrated that when constraints on the velocity of
the gravitational waves are imposed, quantities with different
origins in the action, become interconnected. Specifically, the
scalar potential is not freely chosen but is derived from a
differential equation, once the scalar functions coupled to the
Ricci scalar and the Gauss-Bonnet invariant are specified. The
choice of the scalar potential as a function, which is not freely
specified is not mandatory, but it sure is helpful as it was
demonstrated in the previous sections. Continuing, we demonstrated
that functions which have appealing characteristics, such as the
exponential function, are excellent candidates for describing the
inflationary era as the ratios of the derivatives of the coupling
functions, which appear in the equations of motion, are greatly
simplified. However, in order to make the system of the
gravitational equations solvable, certain approximations had to be
made. The choice of approximations may vary as there exist a lot
possible configurations of Hubble's parameter $H^2$ and its
derivatives $\dot H$, so this approach is extremely model
dependent. As a result, the equations which are derived are fully
solvable in an analytic way, and in certain cases they are also
elegant. Moreover, many models which could seem to be able
to manifest compatible results with the recent observations may be
not valid models in fact, due to the violation of even a single
approximation, so one must be very careful when working with any
model and following either approach. Finally, we presented the
different possible approaches and considered the corresponding
approximations which must apply in order for the model to be
rendered viable and easy to solve analytically. We should note
that our formalism can easily be applied for the cases that the
scalar potential is absent, so this would result to a constraint
between the Einstein-Gauss-Bonnet scalar coupling function $\xi
(\phi)$ and the function $h(\phi)$. We aim to address this
interesting subclass of theories in a future work.

\section*{Acknowledgments}

This work is supported by MINECO (Spain), FIS2016-76363-P, and by
project 2017 SGR247 (AGAUR, Catalonia) (S.D.O).


\begin{thebibliography}{99}





\bibitem{Riess:1998cb}
  A.~G.~Riess {\it et al.} [Supernova Search Team],
  Astron.\ J.\  {\bf 116} (1998) 1009
  [astro-ph/9805201].



\bibitem{Nojiri:2017ncd}
S.~Nojiri, S.~D.~Odintsov and V.~K.~Oikonomou,
Phys.\ Rept.\ {\bf 692} (2017) 1 doi:10.1016/j.physrep.2017.06.001
[arXiv:1705.11098 [gr-qc]].

\bibitem{Nojiri:2010wj}
S.~Nojiri and S.~D.~Odintsov,
Phys.\ Rept.\ {\bf 505} (2011) 59
doi:10.1016/j.physrep.2011.04.001 [arXiv:1011.0544 [gr-qc]].

\bibitem{Nojiri:2006ri}
S.~Nojiri and S.~D.~Odintsov,
eConf C {\bf 0602061} (2006) 06
 [Int.\ J.\ Geom.\ Meth.\ Mod.\ Phys.\ {\bf 4} (2007) 115]
doi:10.1142/S0219887807001928 [hep-th/0601213].

\bibitem{Capozziello:2011et}
S.~Capozziello and M.~De Laurentis,
Phys.\ Rept.\ {\bf 509} (2011) 167
doi:10.1016/j.physrep.2011.09.003 [arXiv:1108.6266 [gr-qc]].

\bibitem{Capozziello:2010zz}
V.~Faraoni and S.~Capozziello,
Fundam.\ Theor.\ Phys.\ {\bf 170} (2010).
doi:10.1007/978-94-007-0165-6

\bibitem{delaCruzDombriz:2012xy}
A.~de la Cruz-Dombriz and D.~Saez-Gomez,
Entropy {\bf 14} (2012) 1717 doi:10.3390/e14091717
[arXiv:1207.2663 [gr-qc]].

\bibitem{Olmo:2011uz}
G.~J.~Olmo,
Int.\ J.\ Mod.\ Phys.\ D {\bf 20} (2011) 413
doi:10.1142/S0218271811018925 [arXiv:1101.3864 [gr-qc]].



\bibitem{Ezquiaga:2017ekz}
  J.~M.~Ezquiaga and M.~Zumalacarregui,
  Phys.\ Rev.\ Lett.\  {\bf 119} (2017) no.25,  251304
  doi:10.1103/PhysRevLett.119.251304
  [arXiv:1710.05901 [astro-ph.CO]].









\bibitem{Hwang:2005hb}
  J.~c.~Hwang and H.~Noh,
  Phys.\ Rev.\ D {\bf 71} (2005) 063536
  doi:10.1103/PhysRevD.71.063536
  [gr-qc/0412126].


\bibitem{Nojiri:2006je}
  S.~Nojiri, S.~D.~Odintsov and M.~Sami,
  Phys.\ Rev.\ D {\bf 74} (2006) 046004
  doi:10.1103/PhysRevD.74.046004
  [hep-th/0605039].




\bibitem{Cognola:2006sp}
  G.~Cognola, E.~Elizalde, S.~Nojiri, S.~Odintsov and S.~Zerbini,
  Phys.\ Rev.\ D {\bf 75} (2007) 086002
  doi:10.1103/PhysRevD.75.086002
  [hep-th/0611198].



\bibitem{Nojiri:2005vv}
  S.~Nojiri, S.~D.~Odintsov and M.~Sasaki,
  Phys.\ Rev.\ D {\bf 71} (2005) 123509
  doi:10.1103/PhysRevD.71.123509
  [hep-th/0504052].


\bibitem{Nojiri:2005jg}
  S.~Nojiri and S.~D.~Odintsov,
  Phys.\ Lett.\ B {\bf 631} (2005) 1
  doi:10.1016/j.physletb.2005.10.010
  [hep-th/0508049].







\bibitem{Satoh:2007gn}
  M.~Satoh, S.~Kanno and J.~Soda,
  Phys.\ Rev.\ D {\bf 77} (2008) 023526
  doi:10.1103/PhysRevD.77.023526
  [arXiv:0706.3585 [astro-ph]].



\bibitem{Bamba:2014zoa}
  K.~Bamba, A.~N.~Makarenko, A.~N.~Myagky and S.~D.~Odintsov,
  JCAP {\bf 1504} (2015) 001
  doi:10.1088/1475-7516/2015/04/001
  [arXiv:1411.3852 [hep-th]].


\bibitem{Yi:2018gse}
  Z.~Yi, Y.~Gong and M.~Sabir,
  Phys.\ Rev.\ D {\bf 98} (2018) no.8,  083521
  doi:10.1103/PhysRevD.98.083521
  [arXiv:1804.09116 [gr-qc]].


\bibitem{Guo:2009uk}
  Z.~K.~Guo and D.~J.~Schwarz,
  Phys.\ Rev.\ D {\bf 80} (2009) 063523
  doi:10.1103/PhysRevD.80.063523
  [arXiv:0907.0427 [hep-th]].


\bibitem{Guo:2010jr}
  Z.~K.~Guo and D.~J.~Schwarz,
  Phys.\ Rev.\ D {\bf 81} (2010) 123520
  doi:10.1103/PhysRevD.81.123520
  [arXiv:1001.1897 [hep-th]].


\bibitem{Jiang:2013gza}
  P.~X.~Jiang, J.~W.~Hu and Z.~K.~Guo,
  Phys.\ Rev.\ D {\bf 88} (2013) 123508
  doi:10.1103/PhysRevD.88.123508
  [arXiv:1310.5579 [hep-th]].



\bibitem{Kanti:2015pda}
  P.~Kanti, R.~Gannouji and N.~Dadhich,
  Phys.\ Rev.\ D {\bf 92} (2015) no.4,  041302
  doi:10.1103/PhysRevD.92.041302
  [arXiv:1503.01579 [hep-th]].


\bibitem{vandeBruck:2017voa}
  C.~van de Bruck, K.~Dimopoulos, C.~Longden and C.~Owen,
  arXiv:1707.06839 [astro-ph.CO].



\bibitem{Kanti:1998jd}
  P.~Kanti, J.~Rizos and K.~Tamvakis,
  Phys.\ Rev.\ D {\bf 59} (1999) 083512
  doi:10.1103/PhysRevD.59.083512
  [gr-qc/9806085].


\bibitem{Kawai:1999pw}
  S.~Kawai and J.~Soda,
  Phys.\ Lett.\ B {\bf 460} (1999) 41
  doi:10.1016/S0370-2693(99)00736-4
  [gr-qc/9903017].



\bibitem{Nozari:2017rta}
  K.~Nozari and N.~Rashidi,
  Phys.\ Rev.\ D {\bf 95} (2017) no.12,  123518
  doi:10.1103/PhysRevD.95.123518
  [arXiv:1705.02617 [astro-ph.CO]].



\bibitem{Chakraborty:2018scm}
  S.~Chakraborty, T.~Paul and S.~SenGupta,
  Phys.\ Rev.\ D {\bf 98} (2018) no.8,  083539
  doi:10.1103/PhysRevD.98.083539
  [arXiv:1804.03004 [gr-qc]].



\bibitem{Odintsov:2018zhw}
  S.~D.~Odintsov and V.~K.~Oikonomou,
  Phys.\ Rev.\ D {\bf 98} (2018) no.4,  044039
  doi:10.1103/PhysRevD.98.044039
  [arXiv:1808.05045 [gr-qc]].


  \bibitem{Kawai:1998ab}
  S.~Kawai, M.~a.~Sakagami and J.~Soda,
  Phys.\ Lett.\ B {\bf 437}, 284 (1998)
  doi:10.1016/S0370-2693(98)00925-3
  [gr-qc/9802033].


\bibitem{Yi:2018dhl}
  Z.~Yi and Y.~Gong,
  Universe {\bf 5} (2019) no.9,  200
  doi:10.3390/Universe5090200
  [arXiv:1811.01625 [gr-qc]].


\bibitem{vandeBruck:2016xvt}
  C.~van de Bruck, K.~Dimopoulos and C.~Longden,
  Phys.\ Rev.\ D {\bf 94} (2016) no.2,  023506
  doi:10.1103/PhysRevD.94.023506
  [arXiv:1605.06350 [astro-ph.CO]].


\bibitem{Kleihaus:2019rbg}
  B.~Kleihaus, J.~Kunz and P.~Kanti,
  arXiv:1910.02121 [gr-qc].





\bibitem{Bakopoulos:2019tvc}
  A.~Bakopoulos, P.~Kanti and N.~Pappas,
  Phys.\ Rev.\ D {\bf 101} (2020) no.4,  044026
  doi:10.1103/PhysRevD.101.044026
  [arXiv:1910.14637 [hep-th]].


\bibitem{Maeda:2011zn}
  K.~i.~Maeda, N.~Ohta and R.~Wakebe,
  Eur.\ Phys.\ J.\ C {\bf 72} (2012) 1949
  doi:10.1140/epjc/s10052-012-1949-6
  [arXiv:1111.3251 [hep-th]].

\bibitem{Bakopoulos:2020dfg}
  A.~Bakopoulos, P.~Kanti and N.~Pappas,
  arXiv:2003.02473 [hep-th].


\bibitem{Ai:2020peo}
W.~Ai,
[arXiv:2004.02858 [gr-qc]].


\bibitem{Odintsov:2020sqy}
S.~Odintsov, V.~Oikonomou and F.~Fronimos,
[arXiv:2003.13724 [gr-qc]].


\bibitem{Odintsov:2020zkl}
S.~Odintsov and V.~Oikonomou,
[arXiv:2004.00479 [gr-qc]].

\bibitem{Easther:1996yd}
  R.~Easther and K.~i.~Maeda,
  Phys.\ Rev.\ D {\bf 54} (1996) 7252
  doi:10.1103/PhysRevD.54.7252
  [hep-th/9605173].

\bibitem{Antoniadis:1993jc}
  I.~Antoniadis, J.~Rizos and K.~Tamvakis,
  Nucl.\ Phys.\ B {\bf 415} (1994) 497
  doi:10.1016/0550-3213(94)90120-1
  [hep-th/9305025].


\bibitem{GBM:2017lvd}
  B.~P.~Abbott {\it et al.}
  ``Multi-messenger Observations of a Binary Neutron Star Merger,''
  Astrophys.\ J.\  {\bf 848} (2017) no.2,  L12
  doi:10.3847/2041-8213/aa91c9
  [arXiv:1710.05833 [astro-ph.HE]].






\bibitem{Odintsov:2019clh}
  S.~D.~Odintsov and V.~K.~Oikonomou,
  Phys.\ Lett.\ B {\bf 797} (2019) 134874
  doi:10.1016/j.physletb.2019.134874
  [arXiv:1908.07555 [gr-qc]].








\bibitem{Akrami:2018odb}
  Y.~Akrami {\it et al.} [Planck Collaboration],
  arXiv:1807.06211 [astro-ph.CO].




























\end{thebibliography}
\end{document}